# A compact fine-grained calorimeter for luminosity measurement at a linear collider

I. Bozovic Jelisavcic[a]

[on behalf of the FCAL Collaboration]

[a]Vinca Institute of Nuclear Sciences, University of Belgrade 11000 Belgrade, Serbia

*Abstract:* Based on a paper published in 2019 by the FCAL Collaboration, this talk is giving an update of the Collaboration's effort to design prototype of highly compact calorimeter to instrument the very forward region of a detector at future $e^+e^-$ colliders. A luminometer prototype, based on sub-millimeter thick detector planes, is tested with an electron-beam of energy 1–5 GeV. The effective Molière radius of the prototype comprising eight detector planes was measured to be $(8.1 \pm 0.1 \text{ (stat)} \pm 0.3 \text{ (syst)})$ mm, and the result is well reproduced by the Monte Carlo simulation.

## 1 Introduction

Luminometer for a future electron-positron collider have challenging requirements derived primarily from a relative precision of integrated luminosity foreseen to be at the level of $10^{-3}$-$10^{-4}$ [1]. This requirement translates into a very extensive list of systematic uncertainties, both detector and experimental environment related, to be controlled at the same level of precision as the luminosity. Compactness of the luminometer is required in order to enable precision position and energy determination of Bhabha final states, based on the fact that an electromagnetic shower is almost fully (90%) contained in the calorimeter volume with radial dimension $R_M$. It is called Molière radius, and here is defined as an effective observable for the prototyped structure. It was experimentally determined in the 2016 test-beam campaign at DESY, in an electron-beam of energy 1-5 GeV.

### 1.1 Prototype of a luminometer

The luminometer is designed as a Silicon-Tungsten sandwich calorimeter with 30-40 sensor planes, depending on a collider center-of-mass energy, followed by one radiation length thick ($1X_0 \approx 3.5$mm) absorber planes. Silicon sensors (320 μm thick) are radially segmented into 64 pads with a 1.8 mm pitch and 48 sectors covering 7.5 deg each in the azimuthal angle. Every pad should be read-out by a FLAME ASIC [2] placed outside the calorimeter. In the test-beam, 4-sector sensors were available, read-out with 128 channels with modified APV25 board [2]. Segmentation of the sensor planes is given in Figure 1 (left) [3].

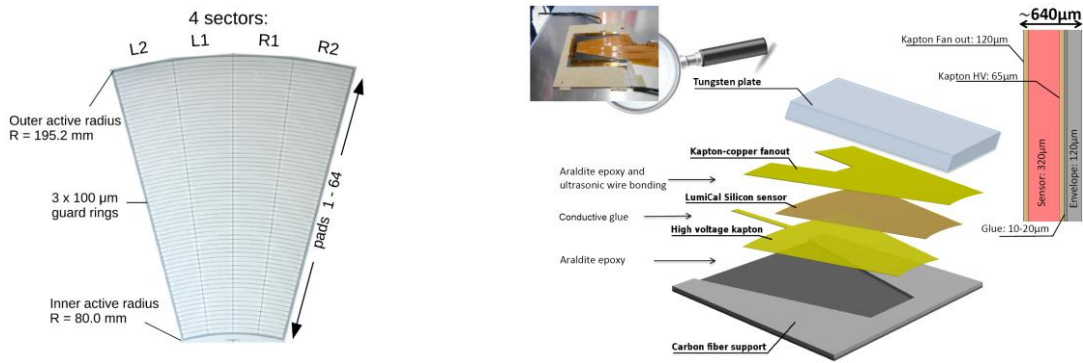

Figure 1: Segmentation of a sensor plane (left). Structure of a detector plane (right).

---

[a] E-mail: ibozovic@vinca.rs

Detector planes are assembled in a way that the sensor is attached to the high-voltage kapton with a conductive glue, while the kapton-copper fan-out is connected with a sensor via ultrasonic wire bonding (50-100 μm loop height). Eight ultra-thin detector planes are assembled, with the overall thickness of 640 μm each [3]. This is illustrated in Figure 1 (right) [3]. Detector planes are placed into 1 mm gaps between tungsten absorbers. Detector and absorber planes are inserted into precision a mechanical structure developed [4]. Image of the overall prototype is given in Figure 2 [3].

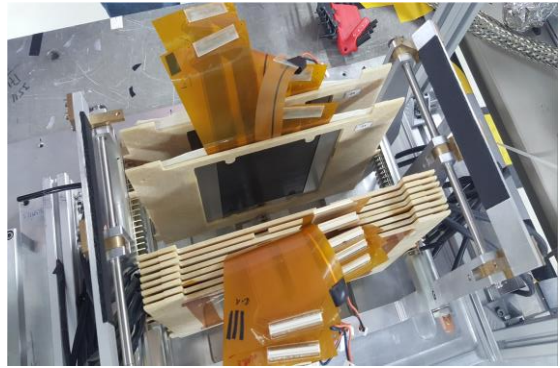

Figure 2: Photograph of the assembled prototype.

## 2 Test-beam setup

As illustrated in Figure [3], two detector planes were used as a tracker while five detector planes and eight absorber layers make the calorimeter prototype of a total thickness of 7.7 $X_0$. Electron beam is passed through a 5x5 mm$^2$ square collimator that limits the beam-spread. The AIDA/EUDET beam telescope was placed upstream of the calorimeter. The telescope was split into two parts T1 and T2, each containing an arm with 3 layers of MIMOSA-26 pixel silicon detectors and 2 thin scintillator counters Sc1 and Sc2, for the trigger system. A dipole magnet of 13 kGs was employed for separation of electrons from photons generated in a copper target that was mounted just in front of the magnet. The beam test aimed to study the performance of the compact calorimeter, and these results are presented in [3], and to test the concept of tracking detectors in front of the calorimeter as a tool for electron and photon identification, where a dedicated data analysis is ongoing.

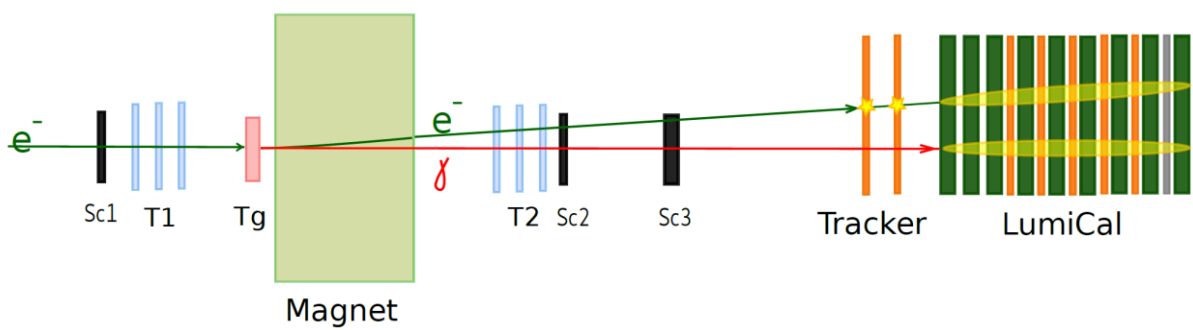

Figure 3: Scheme of the 2016 test-beam setup.

# 3 Results of the test-beam

More than seven million events were collected in the electron test-beam from 1 GeV to 5 GeV energy, with 1 GeV steps, for different setup configurations, to measure the precision of the shower position determination, the electromagnetic shower development in longitudinal and transverse directions and the effective Molière radius. A brief overview is given below.

*3.1 Overall performance*

In Figure 4, linearity of a detector response to the beam-energy is shown. The measured values (red) exhibit a tendency to reduce the slope at larger beam energies. After applying the APV25 calibration and corrections to the leakage fraction estimated from simulation (blue), the dependency is linear.
Front-end electronics' performance is reasonable, with a signal to noise ratio of 7-10 for all channels (Figure 5, left), while the signal efficiency is decreasing for small signals of a few MIPs (Figure 5, right). One of the goals to be achieved with the FLAME ASIC is a larger input range.

*3.2 Measurement of the shower position*

The two tracker planes in front of the calorimeter prototype enable measurement of the radial position resolution of a shower. Reconstructed radial position of a shower is compared w.r.t. the tracker hits and is showing a resolution of (440±20) μm for 5 GeV electrons (Figure 6).

*3.3 Effective Molière radius*

The transverse profile of a shower (Figure 7, left) is fitted with a function $F_E$ (Eq. 1), where a Gaussian term describes a shower core and a modified Grindhammer-Peters term [4] describes the tail of the profile,

$$F_E(r) = A_C e^{-(\frac{r}{R_C})^2} + A_T \frac{2r^\alpha R_T^2}{(r^2 + R_T^2)^2} \quad (1)$$

where $A_C$, $R_C$, $A_T$, $R_T$ and α are free parameters to be determined by fitting $F_E$ to the measured distribution. From Figure 7 (left), one can see that the Geant4 based Monte Carlo simulation matches the experimental data within uncertainties. Integration over r, from 0 to $R_M$, defines 90% of the deposited energy. The effective Molière radius $R_M$ from this requirement is found to be: (8.1±0.1(stat.)±0.3(syst.)) mm, for 5 GeV electrons, proving the feasibility of constructing a compact calorimeter with the proposed design. The effective Molière radius depends a bit on the electron energy due to a limited longitudinal coverage with the existing number of detector planes. From simulation, it is clear that 20 planes should suffice even for 10 GeV electron shower containment (Figure 7, right). The effective Molière radius also (linearly) depends on the air-gap size, that was reduced in the 2016 test-beam w.r.t. the previous campaigns.

Together with maximization of the instrumented sensor area (256 channels instead of 128) with the FALME ASIC, more detector planes are foreseen to be tested in the 2020 test-beam campaigns. In addition to the effective Molière radius measurement and study of electron-photon separation capabilities with a tracker in front of a luminometer, the goal of 2020 test-beam will be to measure energy and polar angle resolution of the detector.

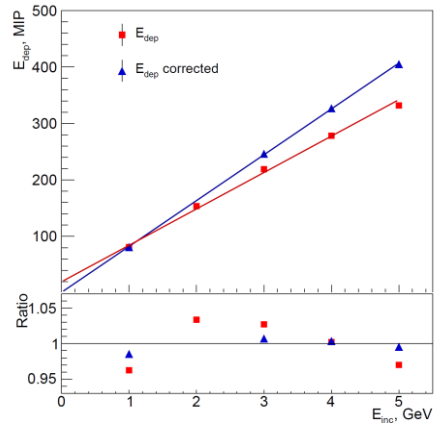

Figure 4: Detector response to the incoming electron beam. The lower part of the figure shows the ratio of deposited energy to the straight line.

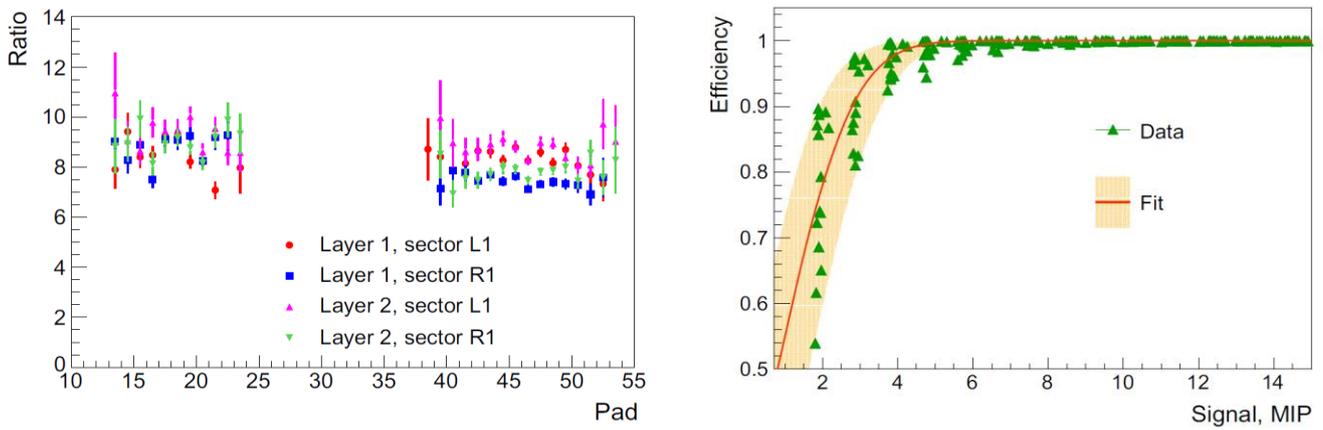

Figure 5: Signal to noise ratio for selected detector planes and sectors (left) and collection efficiency for various signal sizes (right).

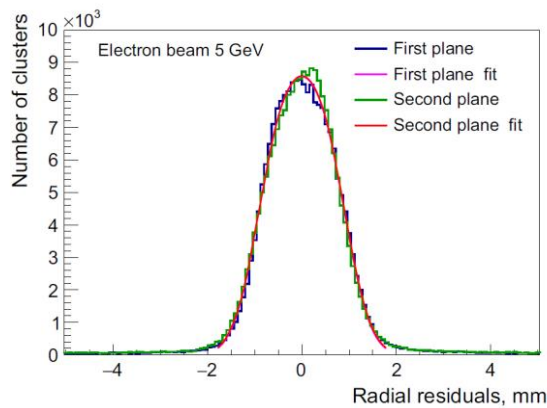

Figure 6: Distribution of shower radial position measurement w.r.t. the tracker hits.

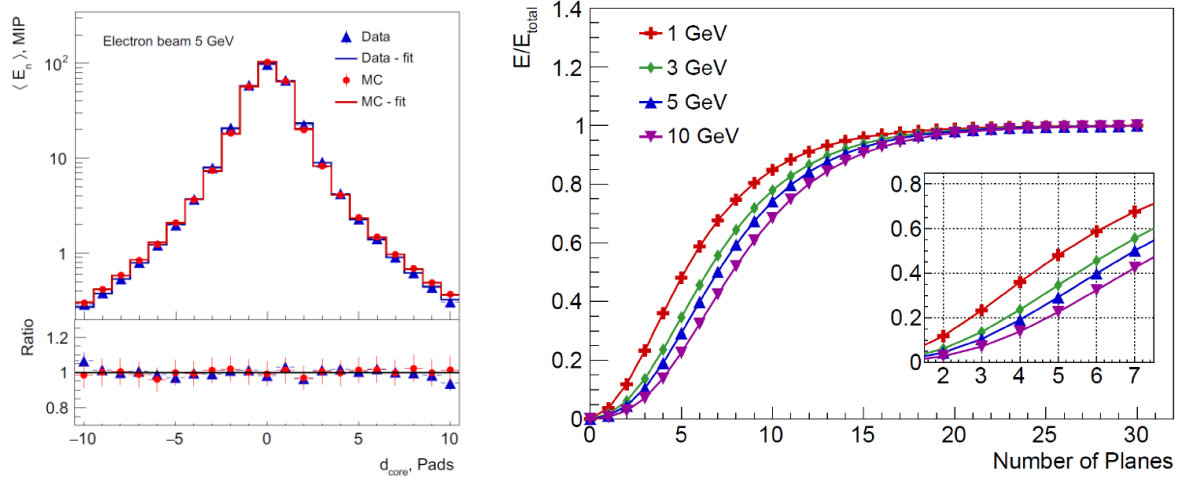

Figure 7: Transverse shower profile for 5 GeV electrons (left) and shower containment dependence on the number of detector planes (right).

## 4 Summary

Compact calorimeters to instrument the very forward region of a detector at an $e^+e^-$ collider are designed, simulated and prototyped by the FCAL Collaboration. Such a calorimeter is consistent with the conceptual design of detectors for ILC and CLIC and optimized for a high precision luminosity measurement. The effective Molière radius $R_M=(8.1\pm0.1(stat.)\pm0.3(syst.))$ mm of the prototyped structure, measured in the test-beam, demonstrates the feasibility of a compact calorimetry. For the first time in this effort, sub-millimeter detector planes are produced. Detector prototype exhibited linearity of response to 1-5 GeV electron test-beam. The measured shower reconstruction precision and shower development are in agreement with the Monte Carlo expectations. Further efforts will lead in the direction of production of the dedicated front-end electronics, with a larger input range and maximization of the instrumented sensor area.